\documentclass[a4paper,11pt]{article}
\pdfoutput=1
\usepackage{jheppub}

\usepackage{amsmath}
\usepackage{amssymb}
\usepackage{amsfonts}
\usepackage{bbm}
\usepackage{braket}
\usepackage{caption}
\usepackage{graphicx}
\usepackage{color}
\usepackage{tabls}
\usepackage{hyperref}
\usepackage{subcaption}
\usepackage{float}
\usepackage{tikz}
\usepackage{orcidlink}
\numberwithin{equation}{section}

\usepackage[numbers,sort&compress]{natbib}
\DeclareMathOperator{\SU}{\mathrm{SU}}
\DeclareMathOperator{\U}{\mathrm{U}}
\DeclareMathOperator{\Z}{\mathbb{Z}}

\newcommand{\SW}{S_{\mbox{\tiny{W}}}}
\newcommand{\SV}{S_{\mbox{\tiny{V}}}}
\newcommand{\Ns}{N_{\mbox{\tiny{s}}}}
\newcommand{\Nt}{N_{\mbox{\tiny{t}}}}
\newcommand{\Ploop}{\mathcal{L}}

\newcommand{\dd}{{\rm{d}}}
\newcommand{\lambdaD}{\lambda_{\mbox{\tiny{D}}}}
\newcommand{\Tc}{T_{\mbox{\tiny{c}}}}
\newcommand{\betac}{\beta_{\mbox{\tiny{c}}}}
\newcommand{\Up}{U_{\mbox{\tiny{p}}}}

\title{\boldmath On the equation of state of $\U(1)$ lattice gauge theory in three dimensions}

\author[a]{Michele~Caselle,\orcidlink{0000-0001-5488-142X}}
\author[a,b]{Alessandro~Mariani,\orcidlink{0000-0002-8830-920X}}
\author[a,c]{Marco~Panero\orcidlink{0000-0001-9477-3749}}
\author[d]{and Antonio~Smecca\orcidlink{0000-0002-8887-5826}}

\affiliation[a]{Physics Department, University of Turin \& INFN, Turin unit, Via Pietro Giuria 1, I-10125 Turin, Italy}
\affiliation[b]{Albert Einstein Center for Fundamental Physics, Institute for Theoretical Physics, University of Bern, Sidlerstrasse 5, CH-3012 Bern, Switzerland}
\affiliation[c]{Department of Physics \& Helsinki Institute of Physics, University of Helsinki, PL 64, FIN-00014 Helsinki, Finland}
\affiliation[d]{Department of Physics, Faculty of Science and Engineering, Swansea University (Singleton Park Campus), Singleton Park, SA2 8PP Swansea, Wales, United Kingdom}

\emailAdd{caselle@to.infn.it}
\emailAdd{a.mariani@unito.it}
\emailAdd{marco.panero@helsinki.fi}
\emailAdd{antonio.smecca@swansea.ac.uk}

\abstract{We study the equation of state of three-dimensional compact $\U(1)$ gauge theory on the lattice by means of numerical simulations, and discuss the implications of our results for the spectrum of the theory, in connection with previous results from the literature. We also compare our findings to the case of non-Abelian gauge theories and comment on the continuum limit.}

\begin{document}

\begin{flushright}
HIP-2025-02/TH
\end{flushright}

\maketitle
\flushbottom

\section{Introduction}
\label{sec:introduction}

Historically, compact $\U(1)$ lattice gauge theory in three spacetime dimensions has served as a useful toy model to investigate some aspects of the confinement phenomenon in gauge theories, and is now a well-understood model, both analytically~\cite{Polyakov:1976fu, Gopfert:1981im, Gopfert:1981er, Banks:1977cc, Muller:1981ia, Aharony:2024ctf} and numerically~\cite{Karliner:1983ab, Wensley:1989ja, Loan:2002ej, Azcoiti:2009md, Caselle:2014eka, Caselle:2016mqu, Athenodorou:2018sab}. Several studies have also been performed at finite temperature~\cite{Coddington:1986jk, Chernodub:2001ws, Borisenko:2008sc, Borisenko:2010qe, Borisenko:2015jea, Caselle:2019khe}. More recently, this theory has been the subject of revived interest, since it was realized that it provides an effective low-energy description for certain condensed-matter systems~\cite{Read:1990zza, Hermele:2004zz, Takashima:2005qh, Fradkin:2013sab, moessner2021topological}. The crucial feature of the theory, from which its confinement properties follow, is the compactness of the gauge group, which leads to the existence of topologically non-trivial field configurations, i.e., instantons.\footnote{In the literature, such configurations are sometimes called ``monopoles'' (or ``monopole-instantons'').}

Despite the amount of knowledge about the theory, there still remain some open issues. For example, a recent, state-of-the-art lattice study~\cite{Athenodorou:2018sab} reported numerical evidence that the spectrum of physical states consists of a tower of equally-spaced energy levels as the continuum limit is approached, and suggested the interpretation of the lightest physical particle as a pseudoscalar ``massive photon'', with all other energy levels in the spectrum being states of two or more such particles. Testing this conjecture by simply evaluating the masses of the states beyond the lightest one is not a numerically trivial task, due to the technical challenges that lattice spectroscopy entails~\cite{Padmanath:2018zqw, Vadacchino:2023vnc}: for stable states, the masses can be extracted from hyperbolic-cosine fits (or from the associated effective-mass fits) of finite-volume Euclidean correlation functions of zero-momentum, gauge-invariant lattice operators with the quantum numbers of the target states. However, the fact that the ``wave functions'' of the physical states of interest are \emph{a priori} unknown implies that, in practice, one has to use a sufficiently large basis of interpolating operators in each channel (possibly defined in terms of smeared and/or blocked lattice fields), evaluating matrices of correlation functions between them and solving the associated generalized eigenvalue problem~\cite{Michael:1985ne, Blossier:2009kd}. This procedure can be particularly challenging for heavy states, whose Euclidean correlators decay over just a few units of the typical lattice spacings used in the simulations.

A complementary strategy to probe the spectrum of a theory consists in studying its equation of state in the canonical ensemble. This idea has fruitful applications in the context of nuclear physics, where the equation of state of hadronic matter can be described in the hadron-resonance picture~\cite{Hagedorn:1965st, Hagedorn:1980kb, Hagedorn:1984hz, Fiore:1984yu, Cleymans:1992zc} (see also ref.~\cite{Wheaton:2004qb}): equilibrium thermodynamic quantities can be modelled in terms of a gas of the physical states appearing in the spectrum, with their own multiplicities, and the dependence of the pressure on the temperature $T$ can thus help reveal non-trivial features of the particle content of the model. For purely gluonic non-Abelian gauge theories in four-dimensional spacetime, whose spectrum consists of an exponentially increasing number of glueball states, in the confining phase the hadron-resonance model predicts a characteristic temperature dependence of the pressure: the latter is indeed observed in lattice calculations, including the expected spin and charge-conjugation degeneracies (as well as the trivial color multiplicity, confirming that the states contributing to the thermodynamics are, indeed, confining ones)~\cite{Meyer:2009tq, Caselle:2011fy, Borsanyi:2012ve, Caselle:2015tza}.

For the compact $\U(1)$ theory in three dimensions, the radically different structure of the spectrum\footnote{One may even question the existence of ``glueball-like'' states in an Abelian gauge theory, given that photons carry no electric charge, and this is true also for the continuum theory in three spacetime dimensions. However, this argument does not necessarily apply for the lattice theory at finite spacing, where non-trivial interaction terms among the gauge fields do exist.} is expected to induce a quantitatively different form of the temperature-dependence of the pressure in the confining phase. This observation is related to another interesting feature of the $\U(1)$ model in three dimensions: even though, as we mentioned, it is an interesting and analytically tractable example of a confining gauge theory, its properties are quite different from those of non-Abelian gauge theories (including quantum chromodynamics). This is reflected in the fact that confining flux tubes between opposite charges are described by a different low-energy dynamics, including, in particular, terms that are absent in the effective description of ordinary non-Abelian theories~\cite{Aharony:2024ctf} (for a general discussion on the effective string theory approach to confinement see, for instance, refs.~\cite{Aharony:2013ipa, Caselle:2021eir}).

Similarly, one also expects differences at temperatures above the deconfinement one, $\Tc$. In the $T \gtrsim \Tc$ regime, non-Abelian gauge theories exhibit thermodynamic quantities scaling with the color and spin multiplicities of gluon-like quasi-particles, but clearly inconsistent with the picture of a gas of non-interacting massless particles that is expected in the Stefan--Boltzmann limit~\cite{Boyd:1996bx, Panero:2009tv, Caselle:2011mn, Borsanyi:2012ve, Bruno:2014rxa, Kitazawa:2016dsl, Giusti:2016iqr, Caselle:2018kap, Bruno:2024dha}, due to the existence of non-trivial dynamical effects~\cite{Gross:1980br}---including, in particular, both chromoelectric and chromomagnetic screening. This can be contrasted with the Abelian case, where magnetic screening is absent and where one may argue that the theory would saturate the Stefan--Boltzmann limit immediately above the deconfinement temperature.

To study these issues at the quantitative level, in this work we present a Monte~Carlo determination of the equation of state of three-dimensional $\U(1)$ lattice gauge theory, comparing it with different models in the low- and high-temperature phases. As will be shown below, in the confining phase the equation of state can be described well in terms of a single physical state (consistent with the conjecture formulated in ref.~\cite{Athenodorou:2018sab}), while in the deconfined phase it can be modelled by means of the single transverse degree of freedom associated with a free photon propagating in two spatial dimensions. In passing, we also discuss the inequivalent continuum limits that one can take in this theory, their implications at finite temperature, and comment on analogies and differences with respect to the four-dimensional counterpart of this theory.

The rest of the work is organized as follows. In section~\ref{sec:generalities_definitions_and_lattice_setup} we set the definitions used in the rest of the manuscript, describing the general features of three-dimensional $\U(1)$ lattice gauge theory, and the techniques to compute its equation of state. In section~\ref{sec:results} we present our simulation results, while in section~\ref{sec:discussion_and_conclusions} we comment on their interpretation and summarize our findings.

\section{Generalities, definitions and lattice setup}
\label{sec:generalities_definitions_and_lattice_setup}

In a continuous Minkowski spacetime with two space dimensions, Maxwell's theory can be defined in terms of the action
\begin{align}
\label{continuum_Minkowski_action}
S_{\mbox{\tiny{cont}}} = -\frac{1}{4} \int \dd^2x \int \dd t F_{\mu\nu}F^{\mu\nu},
\end{align}
where $F_{\mu\nu}=\partial_\mu A_\nu - \partial_\nu A_\mu$ denotes the field strength and $A$ is the gauge field. Note that in two space dimensions both the gauge field and the electric charge $e$ have dimension $1/2$, the Coulomb potential is logarithmic, and the ``magnetic'' field is actually a scalar. At the classical level, the fields obey the equations of motion $\partial_\mu F^{\mu\nu}=0$, while the Bianchi identity $\epsilon_{\mu\nu\rho} \partial^\mu F^{\nu\rho}=0$ is trivially satisfied, which allows one to rewrite the theory in terms of a free, massless scalar field $\phi$, such that $\partial_\mu \phi = \epsilon_{\mu\nu\rho} F^{\nu\rho}$.

Upon second quantization, a regularization is required; in this work we focus on the non-perturbative, gauge-invariant regularization of (the Wick-rotated formulation of) the theory on a regular cubic lattice $\Lambda$ of spacing $a$, whose fundamental degrees of freedom are complex phases $U_\mu(x)$ defined on each oriented link between pairs of nearest-neighbor lattice sites $x$ and $x+a\hat{\mu}$ (with $\hat{\mu}$ denoting a unit vector along the positive $\mu$ direction). We take the Euclidean action of the lattice theory to be~\cite{Wilson:1974sk}
\begin{align}
\label{Wilson_action}
\SW = -\frac{1}{a e^2} \sum_{x \in \Lambda} \sum_{1 \le \mu < \nu \le 3} \mathrm{Re} \, \left(U_{\mu\nu}(x)-1\right),
\end{align}
which is defined as a sum over plaquettes $U_{\mu\nu}(x) = U_\mu(x) U_\nu(x+a\hat{\mu})U_\mu^\star(x+a\hat{\nu})U_\nu^\star(x)$. For later convenience, we also introduce $\beta=1/(a e^2)$. In the $a\to 0$ limit, eq.~\eqref{Wilson_action} reduces to eq.~\eqref{continuum_Minkowski_action} if one assumes the relation between the $U_\mu(x)$ variables and the continuum gauge fields to be
\begin{align}
\label{link_definition}
U_\mu(x)=\exp\left[ i e a A_\mu \left(x+\frac{a}{2}\hat{\mu}\right) \right]
\end{align}
and expands it in powers of the lattice spacing. Equation~\eqref{link_definition} makes it clear that the lattice theory defined by eq.~\eqref{Wilson_action} is invariant under $A_\mu \to A_\mu + 2\pi k/(ea)$ for every $k \in \Z$, namely, that in this formulation the gauge group is taken to be compact. As we mentioned above, the gauge-field periodicity has important implications, since it accounts for the existence of instantons. Their condensation in the ground state of the theory leads to linear confinement of electric charges~\cite{Polyakov:1976fu}.

While eq.~\eqref{Wilson_action} defines a compact formulation of $\U(1)$ lattice gauge theory, one can alternatively consider a non-compact formulation where the Boltzmann factor associated with each plaquette is taken to be~\cite{Villain:1974ir}
\begin{align}
\label{Boltzmann_weight_one_plaquette_Villain_action}
\exp\left( -\SV \right) = 
\sum_{n \in \Z} \exp\left[ -\frac{\beta}{2} (\theta_{\mu\nu}(x)-2\pi n)^2 \right]
,
\end{align}
having defined
\begin{align}
\theta_{\mu\nu}(x)=
A_\mu \left(x+\frac{a}{2}\hat{\mu}\right)
+A_\nu \left(x+a\hat{\mu}+\frac{a}{2}\hat{\nu}\right)
-A_\mu \left(x+a\hat{\nu}+\frac{a}{2}\hat{\mu}\right)
-A_\nu \left(x+\frac{a}{2}\hat{\nu}\right).
\end{align}
While they differ at finite lattice spacing, the lattice theories defined by eq.~\eqref{Wilson_action} and by eq.~\eqref{Boltzmann_weight_one_plaquette_Villain_action} share the same continuum limit. In the lattice discretization defined by eq.~\eqref{Boltzmann_weight_one_plaquette_Villain_action}, the theory admits an exact reformulation as a spin model~\cite{Banks:1977cc, Savit:1977fw, Glimm:1977gz, Baaquie:1977ey}, which allows one to prove~\cite{Gopfert:1981er} that the theory is linearly confining for every finite value of the coupling $e$; in addition, at large $\beta$ the characteristic Debye screening length $\lambdaD$ of the instanton plasma and the string tension $\sigma$ scale differently:
\begin{align}
\label{semiclassical_alambdaD}
\frac{\lambdaD}{a} \simeq \frac{\exp(c_2 \beta)}{c_1 \sqrt{\beta}}
\end{align}
and
\begin{align}
\label{semiclassical_a2sigma}
\sigma a^2 \simeq \frac{c_3}{\sqrt{\beta}} \exp(-c_2 \beta)
\end{align}
respectively, with $c_1=2\pi\sqrt{2}$, $c_2 = \frac{1}{32\sqrt{6}} B \left(\frac{1}{24},\frac{11}{24}\right) B\left(\frac{5}{24},\frac{7}{24}\right)\simeq 2.49435508719\dots$ (where $B(x,y)$ denotes Euler's integral of the first kind), and $c_3=4\sqrt{2}/\pi$. Note that these relations mean, in particular, that the value of the Debye screening length in units of the lattice spacing diverges faster than the inverse of the square root of the string tension in lattice spacings in the $\beta\to \infty$ limit. This implies that, if one defines the continuum limit as the limit in which $a\to 0$ with $\lambdaD$ fixed, then one obtains a theory with an exponentially divergent string tension; thinking about the physical states of the spectrum as modelled in terms of closed flux tubes (an intuitive and surprisingly successful picture for non-Abelian gauge theories in four spacetime dimensions~\cite{Isgur:1984bm}), this implies that their mass becomes infinite, leading to the decoupling of any possible ``glueball'' state in the continuum limit. Alternatively, if one defines the continuum limit as $a\to 0$ at fixed $\sigma$, then the Debye screening length diverges, so that any pair of electric probe charges would be subject to the purely Coulombic, logarithmic potential at every distance. 

We carried out a set of numerical simulations in the compact formulation of the lattice theory defined by eq.~\eqref{Wilson_action}, on isotropic lattices of sizes $L^2/T$, where $L=a\Ns$ denotes the extent of each of the two spatial sizes and $T=1/(a\Nt)$ denotes the temperature. Periodic boundary conditions are imposed in the three main directions. Our Monte~Carlo calculations are based on ensembles of configurations produced with a combination of heat-bath and overrelaxation updates of the $U_\mu(x)$ link variables. In the infinite-volume limit, the Polyakov loop
\begin{align}
\Ploop\left(\vec{x}\right)= \prod_{t=0}^{\Nt-1}U_0\left(\vec{x},at\hat{0}\right)
\end{align}
is the order parameter for a thermal deconfining transition at a critical temperature $\Tc$. General arguments~\cite{Svetitsky:1982gs} suggest this deconfinement transition to be in the universality class of the Kosterlitz--Thouless transition~\cite{Kosterlitz:1973xp}, and this prediction is indeed supported by numerical studies~\cite{Coddington:1986jk, Chernodub:2001ws, Loan:2002ej, Borisenko:2010qe, Borisenko:2015jea, Caselle:2019khe}, which also found that the critical value of $\beta$ corresponding to the deconfinement temperature approximately scales as $\Nt$. This means that $\Tc$ scales like $e^2$ in the continuum limit, which, in turn, implies that taking the $a\to 0$ limit at fixed $\Tc$ would lead to a divergent Debye screening length and to a vanishing string tension. Conversely, if one defines the continuum limit assuming that a physical quantity like $\lambdaD$ remains constant, then $\Tc$ diverges, i.e., the theory confines at all temperatures.

The equation of state of the theory can be determined using the integral method~\cite{Engels:1990vr}, which relies on the equality between the pressure $p$ and minus the density of free energy $f=F/V$ in the thermodynamic limit,
\begin{align}
p = -\frac{\partial F}{\partial V}\bigg\lvert_{T} = T \frac{\partial }{\partial V} \ln Z \bigg\lvert_{T} = \frac{T}{V} \ln Z
\end{align}
(where $Z$ is the partition function of the theory), and is based on the idea of rewriting $f$ as the integral of its derivative with respect to $\beta$, yielding
\begin{align}
\frac{p(T)}{T^3} = 3 \Nt^3 \int_0^{\beta(T)} \dd \beta^\prime \left[ \langle\Up(T)\rangle-\langle\Up(0)\rangle\right] ,
\end{align}
where $\beta(T)$ denotes the value of $\beta$ corresponding to temperature $T$, $\Up$ is the real part of the plaquettes averaged over the whole lattice, and the $\langle\Up(0)\rangle$ term, evaluated at the same $\beta^\prime$ as $\langle\Up(T)\rangle$, but on a lattice at (approximately) zero temperature, is subtracted to remove the ultraviolet-divergent, non-thermal contributions to $p(T)$.

In the deconfined phase at $T>\Tc$, the theory is expected to reduce to a gas of massless and non-interacting photons, whose pressure is
\begin{align}
\label{continuum_Stefan_Boltzmann}
\frac{p(T)}{T^3} = \frac{\zeta(3)}{2\pi} = 0.191313298016\dots ,
\end{align}
where $\zeta$ denotes the Euler--Riemann zeta function. The corresponding formula on a cubic lattice with finite $\Nt$ reads~\cite{Caselle:2011mn}
\begin{align}
\label{lattice_Stefan_Boltzmann}
\frac{p(T)}{T^3} = \frac{\zeta(3)}{2\pi} \left[1 + \frac{7}{4}\frac{1}{\Nt^2} \frac{\zeta(5)}{\zeta(3)} + \frac{227}{32}\frac{1}{\Nt^4} \frac{\zeta(7)}{\zeta(3)} + \frac{8549}{128}\frac{1}{\Nt^6} \frac{\zeta(9)}{\zeta(3)} + \cdots\right] .
\end{align}

In the confining phase, one can expect to model the thermodynamics of the theory in terms of massive states, as in the hadron-gas model for the equation of state of quantum chromodynamics. For a system defined in two space dimensions, the contribution to the pressure from each degree of freedom of mass $M$ can be written as~\cite{Caselle:2011fy}
\begin{align}
\label{pressure_of_one_massive_dof}
\frac{p(T)}{T^3} = 2 \left( \frac{M}{2\pi T}\right)^{3/2} \sum_{n=1}^\infty \frac{1}{n^{3/2}} K_{3/2}\left(\frac{n M}{T}\right),
\end{align}
where $K_\nu(z)$ denotes the modified Bessel function of the second kind of order $\nu$ and argument $z$.

Note that, if the $\U(1)$ theory had a spectrum qualitatively similar to the one observed in non-Abelian gauge theories (both in three and in four spacetime dimensions)~\cite{Meyer:2009tq, Caselle:2011fy, Borsanyi:2012ve, Caselle:2015tza, Caselle:2010qd}, then one would expect its spectral density at large mass to tend to an exponentially increasing function, i.e., a Hagedorn spectrum~\cite{Hagedorn:1965st, Hagedorn:1968zz}.

\section{Results}
\label{sec:results}

In this section, we present results from our numerical simulations on lattices with $\Nt = 4$, $6$ and $8$ sites in the Euclidean-time direction, for values of the spatial sizes in units of the lattice spacing $\Ns$ up to $80$ for $\Nt=4$ and $6$, and up to $104$ for $\Nt=8$. As an ultra-local quantity, the plaquette exhibits very little dependence on $\Ns$; thus we combined the weighted average of the mean plaquette values from the largest three volumes for each $\Nt$. In the analysis, we also used the high-precision values for the Debye screening length and the string tension reported in ref.~\cite{Athenodorou:2018sab}.

\begin{figure}
\centering
\includegraphics[width=0.7\textwidth]{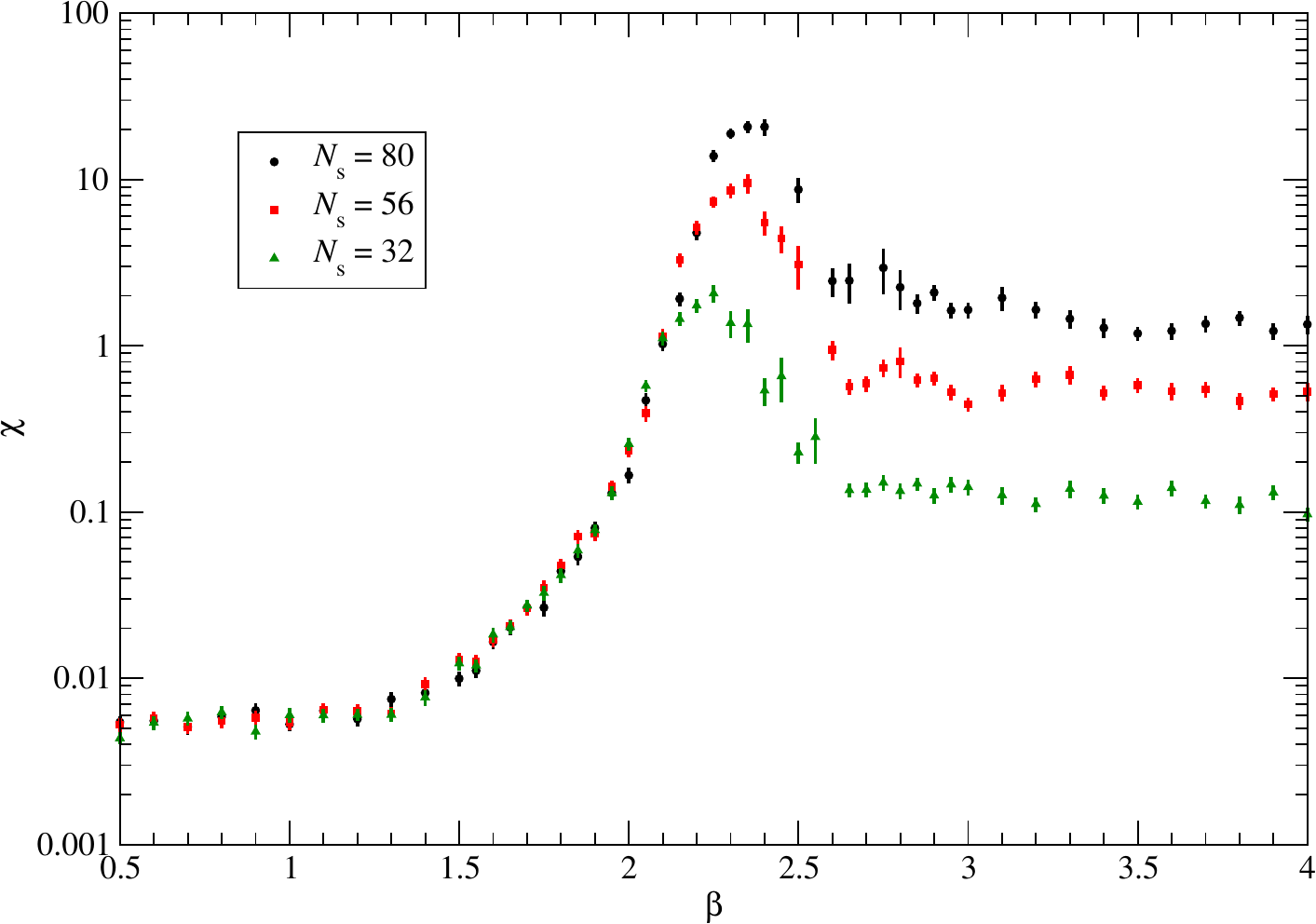}
\caption{Polyakov-loop susceptibility, defined according to eq.~\eqref{susceptibility_definition}, as a function of $\beta$: the figure shows a sample of our results, from simulations on lattices wit temporal extent $\Nt=6$ and spatial extent $\Ns=80$ (black circles), $\Ns=56$ (red squares), and $\Ns=32$ (green triangles).}
\label{fig:Polyakov_susceptibility_Nt_6}
\end{figure}

We estimated the value of the critical coupling $\betac$ for each $\Nt$ as the $\beta$ value corresponding to the peak in the susceptibility $\chi$ associated with (the modulus of) the spatial average of the Polyakov loop
\begin{align}
\label{susceptibility_definition}
\chi = \langle \left| \Ploop \right|^2 \rangle - \langle \left|\Ploop\right| \rangle^2
\end{align}
for the largest volume. In particular, the position of the maximum of $\chi$ was obtained from a fit to a Gau{\ss}ian distribution of the points near the peak of $\chi$; we did not attempt to carry out an extrapolation to the infinite-volume limit. An example of a plot obtained in this part of the analysis is shown in figure~\ref{fig:Polyakov_susceptibility_Nt_6}, for the simulations on lattices with $\Nt=6$.
The $\betac$ values thus obtained are listed in table~\ref{tab:betac}. Such values can be considered as an estimate for the $\beta$ values corresponding to the deconfinement temperature on lattices with $\Nt$ sites in the Euclidean-time direction.

\begin{table}
\centering
\begin{tabular}{ |c|c|c| } 
\hline
$\Nt$ & $\Ns^{\mbox{\tiny{max}}}$ & $\betac$  \\
\hline
$4$   &  $80$                     & $2.11(1)$ \\ 
$6$   &  $80$                     & $2.35(5)$ \\ 
$8$   & $104$                     & $2.55(1)$ \\ 
\hline
\end{tabular}
\caption{Values of the critical coupling $\betac$ corresponding to the peak of the Polyakov-loop susceptibility for the largest spatial size considered (shown in the second column).}
\label{tab:betac}
\end{table}

\begin{figure}
\centering
\includegraphics[width=0.7\textwidth]{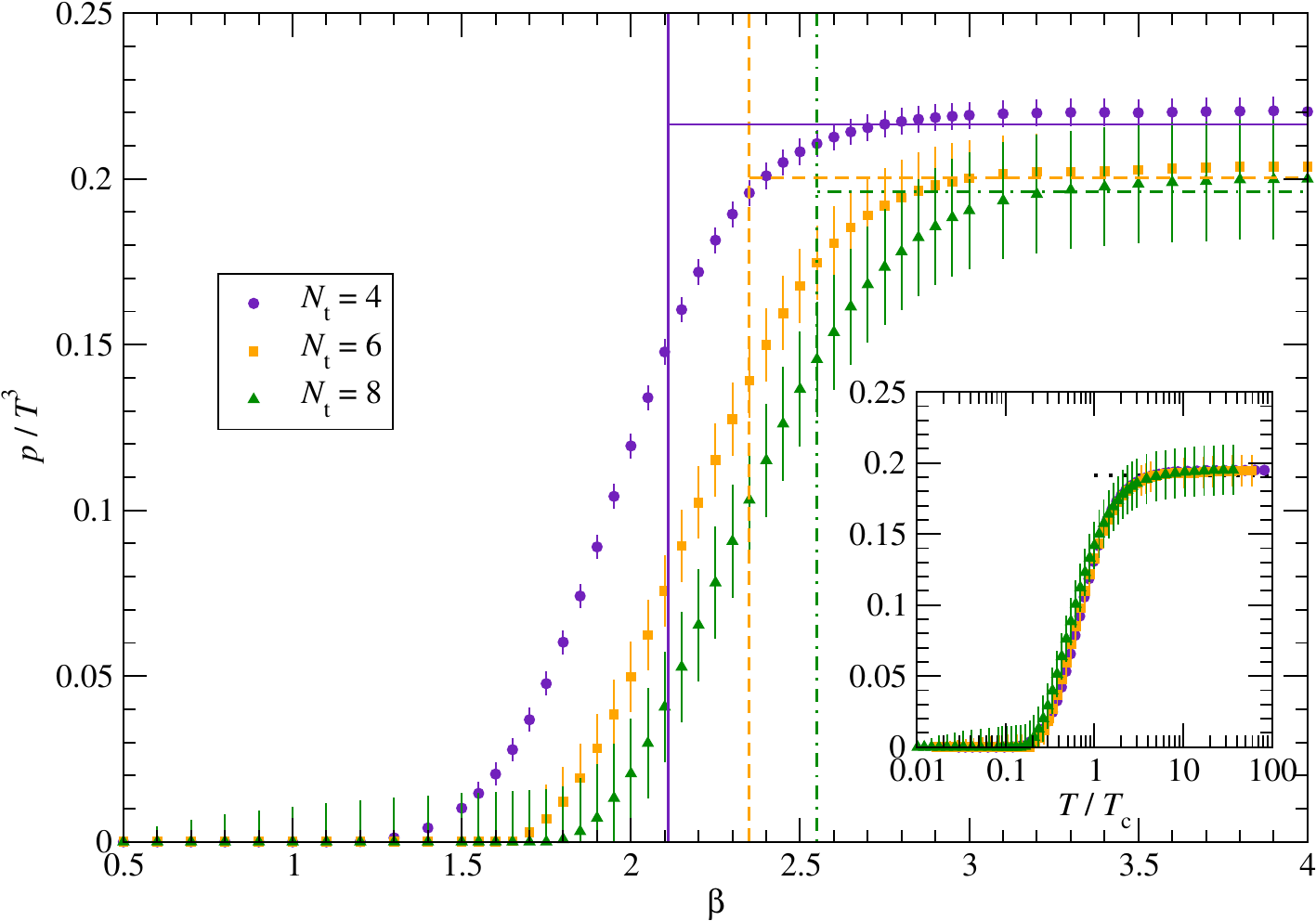}
\caption{Results for the pressure, in units of the third power of the temperature, obtained from simulations on lattices with $\Nt=4$ (indigo circles), $6$ (orange squares), and $8$ (green triangles) sites in the Euclidean-time direction. Data are shown as a function of $\beta=1/(ae^2)$ in the main plot. The vertical straight lines denote the $\betac$ values, corresponding to the maximum of the Polyakov-loop susceptibility, and listed in table~\ref{tab:betac}, while the horizontal lines show the value of the $p/T^3$ ratio predicted for a gas of massless, non-interacting degrees of freedom on a finite-$\Nt$ lattice, according to eq.~\eqref{lattice_Stefan_Boltzmann}: the values corresponding to  $\Nt=4$, $\Nt=6$, and $\Nt=8$ are respectively shown by the solid, dashed, and dash-dotted lines, with the same colors as the data points. The inset plot shows the corresponding results expressed in units of $T/\Tc$, defined in terms of the $\betac$ values yielding the peak of the Polyakov-loop susceptibility, listed in table~\ref{tab:betac}, and rescaled by the factor in square brackets on the right-hand side of eq.~\eqref{lattice_Stefan_Boltzmann}. The dotted horizontal black line denotes the value of the continuum Stefan--Boltzmann limit, eq.~\eqref{continuum_Stefan_Boltzmann}.}
\label{fig:pressure_curve_comparison}
\end{figure}

Our results for the pressure, expressed in units of the third power of the temperature, from simulations on lattices with $\Nt=4$, $6$, and $8$, are shown in the main plot of fig.~\ref{fig:pressure_curve_comparison}, where they are displayed as a function of $\beta$. One can observe that the $p(T)/T^3$ ratio is very small for $\beta \ll \betac$ (namely for $T\ll \Tc$), while it grows quickly for temperatures close to the deconfinement one ($\beta \simeq \betac$), and saturates to the continuum Stefan--Boltzmann value, given in eq.~\eqref{continuum_Stefan_Boltzmann}, within our uncertainties, shortly thereafter. The latter feature is a remarkable difference with respect to the behavior of the equation of state in non-Abelian gauge theories, in which the Stefan--Boltzmann limit is approached very slowly. This behavior of non-Abelian gauge theories can be related to the existence of chromomagnetic screening. In particular, in four spacetime dimensions, the energy scale characterizing chromomagnetic screening is parametrically of the form $g^2T$ at the leading order in perturbation theory, with $g$ denoting the coupling of the non-Abelian gauge theory, and the physical coupling runs logarithmically slowly as a function of the momentum scale at which it is evaluated; in a deconfined Yang--Mills theory, the latter can be identified with the scale of the Matsubara modes, which is of the order of the temperature. Obviously, this description does not apply to an Abelian gauge theory in three spacetime dimensions, whose thermodynamics in the deconfined phase is expected to be drastically different. Our results do confirm this expectation, revealing that the equation of state is consistent with a gas of free photons essentially for all temperatures $T\gtrsim 1.2 \Tc$. This is manifest in the inset plot of fig.~\ref{fig:pressure_curve_comparison}, which shows the results for $p/T^3$ rescaled by the factor in square brackets appearing on the right-hand-side of eq.~\eqref{lattice_Stefan_Boltzmann}, and plotted against $T/\Tc$, where we defined the critical temperature as $\Tc=1/\left(\Nt a(\betac)\right)$, for the $\betac$ values reported in table~\ref{tab:betac}, and using the relation between the lattice spacing and $\beta$ given in ref.~\cite[eq.~(4.5)]{Athenodorou:2018sab} (where $m_{0^{--}}=1/\lambdaD$).

\begin{figure}
\centering
\includegraphics[width=0.7\textwidth]{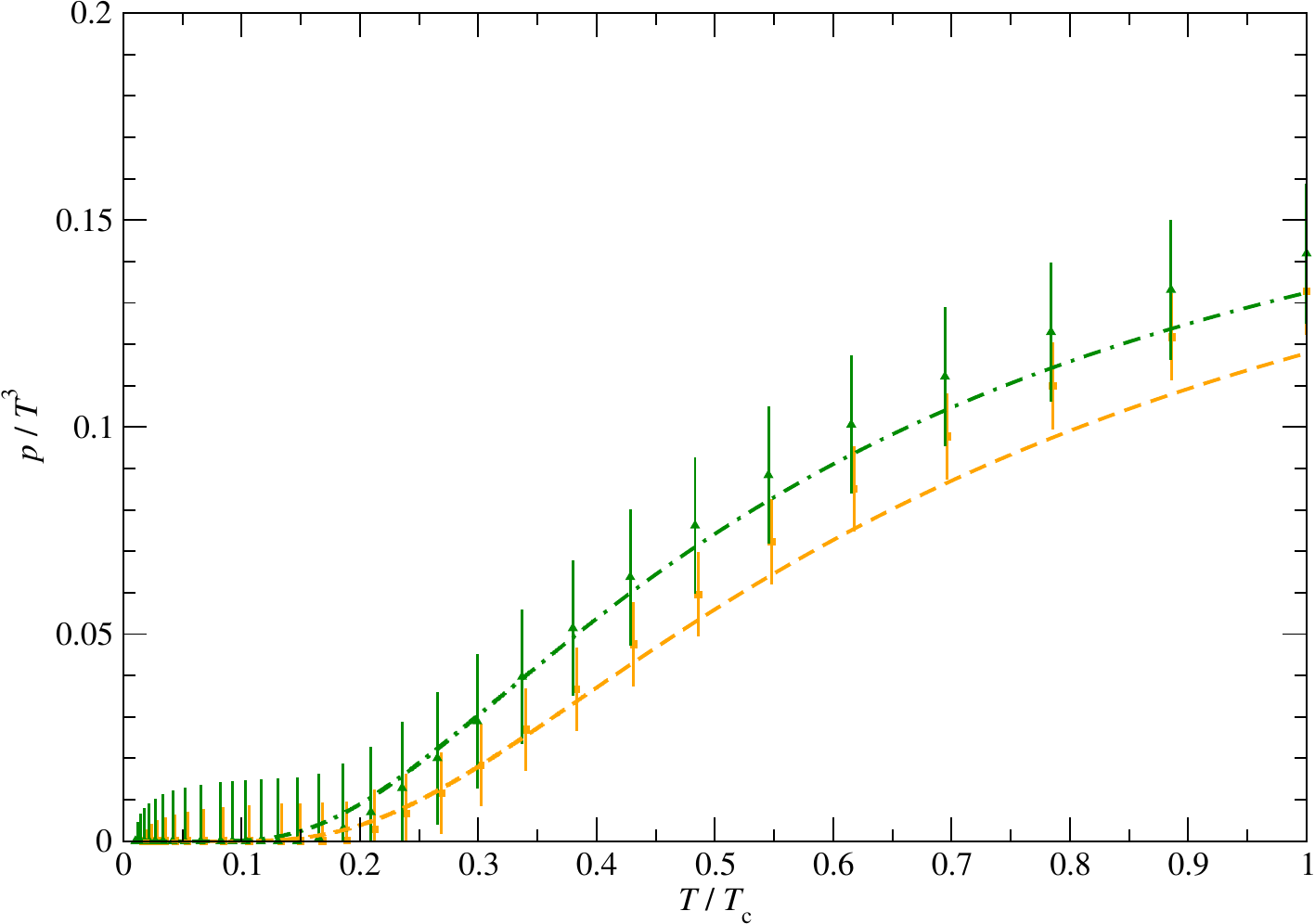}
\caption{Same as in the inset plot of figure~\ref{fig:pressure_curve_comparison}, but for the results from simulations on lattices with $\Nt=6$ (orange squares) and $\Nt=8$ (green triangles) in the confining phase only, in comparison with the values of $p/T^3$ (respectively denoted by the dashed and by the dash-dotted curves) that one would obtain by modelling the thermodynamics of the theory in terms of a gas of states with the single lightest mass determined in ref.~\cite{Athenodorou:2018sab}, according to eq.~\eqref{pressure_of_one_massive_dof}.}
\label{fig:zoom_on_confining_phase}
\end{figure}

While different ways to set the scale are inequivalent in this theory, the choice based on the assumption that the lightest ``mass'' in the spectrum has a physical meaning is particularly convenient to make a comparison of the numerical results for the pressure with a ``hadron-gas-like'' model at $T<\Tc$. As a matter of fact, it is especially interesting to consider the equation of state in the confining phase, as shown in figure~\ref{fig:zoom_on_confining_phase}: if the thermodynamics of the theory were similar to that of non-Abelian gauge theories, then one would expect the $p/T^3$ ratio to approximately follow the dashed curve shown in the plot, which represents the contribution from the lightest glueball-like state determined in ref.~\cite{Athenodorou:2018sab}, only at sufficiently low temperatures. At higher temperatures one should include the contributions from heavier states, too, and, furthermore, sufficiently close to $\Tc$, one would expect an even steeper growth of $p/T^3$ as a function of $T$, which could be interpreted in terms of a Hagedorn spectrum. The results plotted in figure~\ref{fig:zoom_on_confining_phase} show that the behavior of the equation of state of the $\U(1)$ gauge theory is quantitatively very different from this expectation: instead, our simulation results for $p/T^3$ are consistent with the contribution (shown by the solid line) from only the lightest state determined in ref.~\cite{Athenodorou:2018sab}. Clearly, this disproves the existence of a Hagedorn-like spectral density of states and questions the existence of genuinely independent states in the spectrum of this theory.

\begin{figure}
\centering
\includegraphics[width=0.7\textwidth]{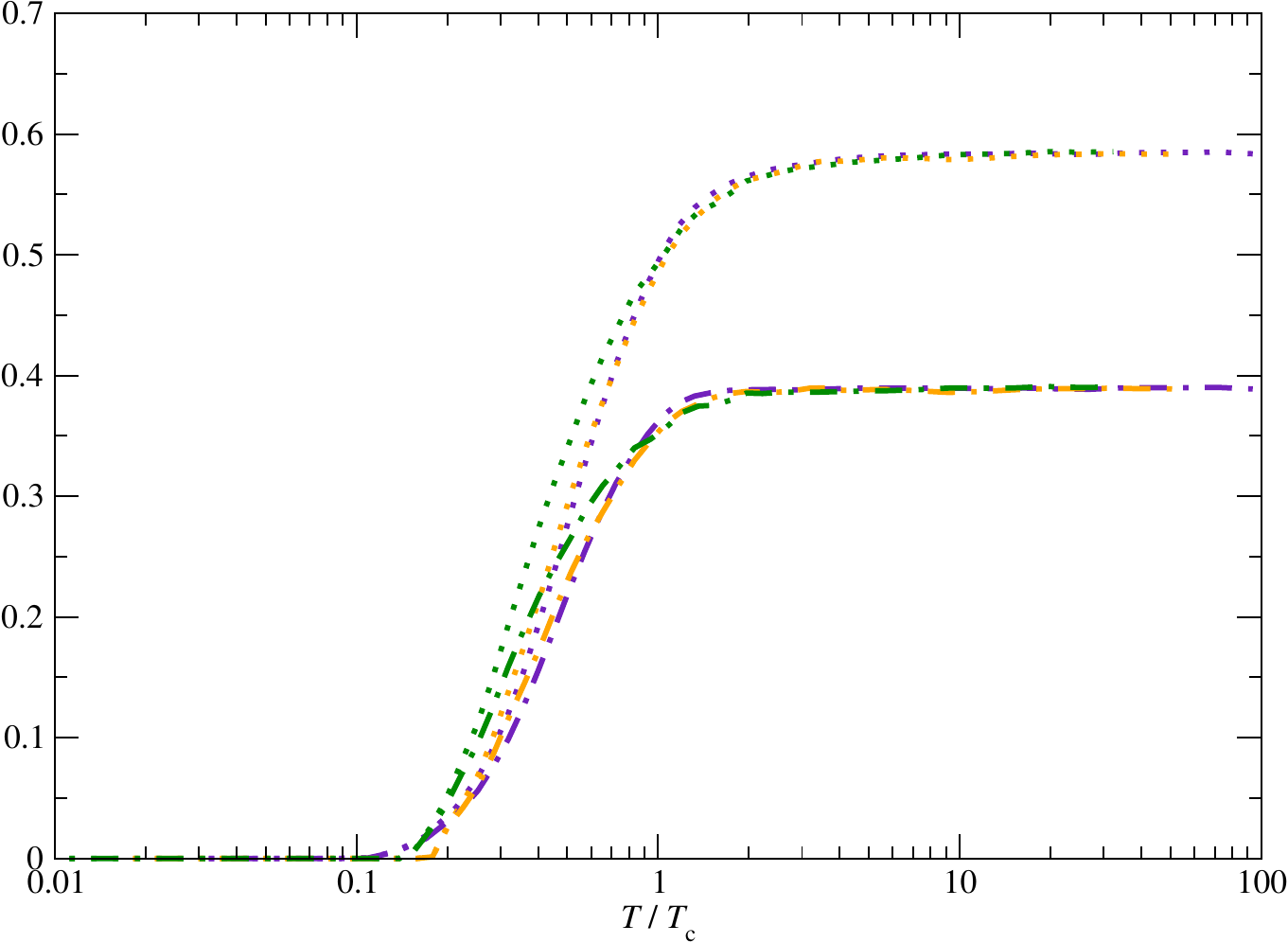}
\caption{Results for the entropy density in units of $T^2$ (dotted lines), and for the energy density, in units of $T^3$ (dash-dotted lines), plotted against $T/\Tc$. Results obtained from simulations on lattices with $\Nt=4$, $\Nt=6$, and $\Nt=8$ are respectively denoted by indigo, orange, and green symbols.}
\label{fig:entropy_energy}
\end{figure}

Finally, the main plot of figure~\ref{fig:entropy_energy} shows our results for other thermodynamic quantities closely related to the pressure, namely the entropy density
\begin{align}
\label{entropy_density}
s=3\frac{p}{T}+T^3\frac{d}{dT}\left(\frac{p}{T^3}\right)
\end{align}
and the energy density
\begin{align}
\label{energy_density}
\epsilon=2p+T^4\frac{d}{dT}\left(\frac{p}{T^3}\right),
\end{align}
expressed in units of the appropriate powers of the temperature and plotted against $T/\Tc$. As for the pressure, the fast approach to the Stefan--Boltzmann limit is manifest, and the essentially trivial behavior of these quantities for almost all temperatures above $\Tc$ can be compared and contrasted with non-Abelian gauge theories.

\section{Discussion and conclusions}
\label{sec:discussion_and_conclusions}

The results reported in section~\ref{sec:results} provide a clear and compelling understanding of the spectral features of the $\U(1)$ theory in three spacetime dimensions, and highlight its remarkable qualitative differences with respect to non-Abelian gauge theories.

On the one hand, the equation of state in the confining phase can be modelled in terms of a single degree of freedom, characterized by a mass that is consistent with the lightest mass determined in ref.~\cite{Athenodorou:2018sab}. This indicates that no contribution from heavier ``would-be glueballs'' is present, and suggests that the spectrum of this theory does not contain the rich variety of states characterizing non-Abelian Yang--Mills theories. Similarly, our results show no evidence of the sharp increase in the pressure at temperatures approaching $\Tc$ from below, which in $\SU(N)$ gauge theories can be interpreted in terms of an exponentially increasing, Hagedorn-like, density of states in the spectrum.

As for the high-temperature phase of the $\U(1)$ theory, we found that the equilibrium thermodynamic quantities already reach the Stefan--Boltzmann limit at temperatures just above $\Tc$, and thus can be modelled in terms of a single non-interacting degree of freedom---the transverse polarization of a photon. Our results also show that, again, there is no evidence of the non-trivial and intrinsically non-perturbative features characteristic of the deconfined phase of $\SU(N)$ gauge theories~\cite{Gross:1980br, Linde:1980ts, Appelquist:1981vg, Braaten:1989mz}. As we already mentioned in section~\ref{sec:results}, this has an obvious explanation: the absence of magnetic screening and the electric neutrality of photons (with the absence of charged matter fields) constrain the thermodynamics in the deconfined phase of the $\U(1)$ theory to be essentially trivial.

It is also interesting to discuss our results from the point of view of (the various ways to take) the continuum limit. In particular, as discussed in section~\ref{sec:generalities_definitions_and_lattice_setup}, the fact that the confining phase of this lattice theory is characterized by two physical length scales (the inverse of the square root of the string tension and the Debye length) that have a different dependence on the coupling, suggests that different continuum limits could exist. However, in none of them does a phase with a linearly confining potential survive: either because the length scale at which the linear potential sets in diverges, or because the string tension itself diverges, or it is impossible to add charged probed sources or charged particles in the continuum theory; at the same time, in none of the different continuum limits can glueballs exist~\cite{Athenodorou:2018sab}.

To summarize, our present study of the equation of state elucidated some aspects about the spectrum of the $\U(1)$ gauge theory in three dimensions, in its compact formulation on the lattice. Part of our motivation for this study stemmed from the fact that, even though this model is well understood, in the literature there remain some claims whose interpretation appeared, at least to us, slightly ambiguous and/or potentially deceptive. A facet of the problem is that, as we discussed in section~\ref{sec:generalities_definitions_and_lattice_setup}, intuition led by properties that are familiar in $\SU(N)$ gauge theories in four spacetime dimensions---including, in particular, a rich spectrum of glueballs, in which the lightest particles dominate the pressure of the theory at low temperatures, while at higher temperatures an exponentially increasing density of states results into a faster growth of equilibrium-thermodynamics quantities as the deconfinement temperature is approached---is misleading, when applied to the $\U(1)$ theory in three dimensions.

In principle, the determination of the critical temperature $\Tc$ that we relied upon could be refined by carrying out a dedicated finite-size-scaling analysis, while in the present work we limited ourselves to estimate the critical values of $\beta$ from the location of the Polyakov-loop susceptibility peak on the largest lattices that we simulated, but without an extrapolation to the thermodynamic limit. While a systematic finite-size-scaling study could yield a more precise determination of $\Tc$, we do not expect that its results could modify the findings of the present work at a qualitative level. Concerning this aspect, it is nevertheless worth remarking that the authors of ref.~\cite{Borisenko:2015jea} pointed out that the extrapolation of $\betac$ to the infinite-volume limit is non-trivial.\footnote{We thank O.~Borisenko and A.~Papa for their comments on this issue.} In particular, the careful investigation carried out in that work (through simulations in the dual formulation, on lattices with $\Nt=8$ and for $\Ns$ values up to $\Ns=512$, using the second-moment correlation length to study the transition) led to an estimate of the pseudocritical $\beta$ around $\betac \simeq 5.6$, i.e., significantly larger than their previous estimate from ref.~\cite{Borisenko:2010qe}. At the same time, they also pointed out that the critical index $\eta$ determined in ref.~\cite{Borisenko:2015jea} is consistent with the value expected from the Svetitsky--Yaffe conjecture~\cite{Svetitsky:1982gs}, namely $\eta=1/4$~\cite{Kosterlitz:1973xp}, whereas that was not the case for the results obtained from simulations on smaller lattices reported in ref.~\cite{Borisenko:2010qe}. These observations indicate that the theory has a logarithmically slow convergence to the thermodynamic limit, and can be compared with similar findings that were reported for the XY model in two dimensions in ref.~\cite{Hasenbusch:2005xm}. For larger and larger values of $\Nt$, one can thus expect the computational costs of a full-fledged thermodynamic-limit analysis to quickly become prohibitive.

As for possible generalizations of the present work, one could carry out a study similar to the one discussed herein for compact $\U(1)$ lattice gauge theory in four spacetime dimensions: while an exact solution of this theory is not known, there exist both analytical~\cite{Guth:1979gz, Frohlich:1982gf, Polyakov:1975rs, Polyakov:1976fu, Banks:1977cc} and numerical studies (see, e.g., refs.~\cite{Zach:1995ni, Zach:1997yz, Vettorazzo:2004cr, Koma:2003gi, Panero:2004zq, Panero:2005iu, Berg:2006hh, Tagliacozzo:2006gu, Heinzl:2007kx, Amado:2013rja} and older works mentioned therein). An important aspect of this theory, though, is that the confining phase at strong coupling and the deconfined phase are separated by a first-order transition, albeit a weak one, making it impossible to properly define a continuum limit. It is perhaps worth mentioning that, while the non-perturbative effects that we studied in the present work can be traced back to the lattice regularization, genuinely physical non-perturbative effects do occur and can be relevant even for Abelian gauge theories that exist in nature, like quantum electrodynamics: a prominent example is given by high-intensity lasers (for a recent review, see ref.~\cite{Fedotov:2022ely}), in which the photon density can significantly exceed one photon per Compton wavelength cubed, and, as a result of coherence, the effective coupling between charged matter fields and photons can become large (even though the coupling itself is small), triggering Schwinger pair-production~\cite{Schwinger:1951nm} and other interesting phenomena of non-perturbative nature.

\section*{Acknowledgments}

We thank O.~Aharony, O.~Borisenko, A.~Bulgarelli, E.~Cellini, A.~Nada and A.~Papa for useful discussions. A.~M. acknowledges funding from the Schweizerischer Nationalfonds (grant agreement number 200020\_200424). The work of A.~S. is supported by the STFC consolidated grant No. ST/X000648/1. We acknowledge support from the SFT Scientific Initiative of INFN. This work was partially supported by the Simons Foundation grant 994300 (``Simons Collaboration on Confinement and QCD Strings''), by the Italian PRIN ``Progetti di Ricerca di Rilevante Interesse Nazionale -- Bando 2022'', prot. 2022ZTPK4E and prot. 2022TJFCYB, and by the Spoke 1 ``FutureHPC \& BigData'' of the Italian Research Centre in High-Performance Computing, Big Data and Quantum Computing (ICSC), funded by the European Union -- NextGenerationEU. The numerical simulations were performed at the ITP cluster of the University of Bern.\\

\noindent {\bf Open Access Statement}---For the purpose of open access, the authors have applied a Creative Commons Attribution (CC BY) licence to any Author Accepted Manuscript version arising.

\bibliography{paper}

\end{document}